# Magnetic resonance study of bulk and thin film EuTiO$_3$

V. V. Laguta[1,2], S. Kamba[1], M. Maryško[1], B. Andrzejewski[3], M. Kachlík[4], K. Maca[4], J. H. Lee[5], D. G. Schlom[6,7]

[1]Institute of Physics ASCR, Na Slovance 2, 18221 Prague 8, Czech Republic
[2]Institute of Physics, Opole University, Oleska 48, 45-052 Opole, Poland
[3]Institute of Molecular Physics PAS, Smoluchowskiego 17, 60-179 Poznań, Poland
[4]CEITEC BUT, Brno University of Technology, Technická 2, 61669 Brno, Czech Republic
[5]Korea Atomic Energy Research Institute (KAERI), Daejeon, Korea
[6]Department of Materials Science and Engineering, Cornell University, Ithaca, New York, 14853-1501, USA
[7]Kavli Institute at Cornell for Nanoscale Science, Ithaca, New York, 14853, USA

**Abstract**

Magnetic resonance spectra of EuTiO$_3$ in both bulk and thin film form were taken at temperatures from 3-350 K and microwave frequencies from 9.2-9.8 and 34 GHz. In the paramagnetic phase, magnetic resonance spectra are determined by magnetic dipole and exchange interactions between Eu$^{2+}$ spins. In the film, a large contribution arises from the demagnetization field. From detailed analysis of the linewidth and its temperature dependence, the parameters of spin-spin interactions were determined: the exchange frequency is 15-15.5 GHz and the estimated critical exponent of the spin correlation length is $\approx 0.5$. In the bulk samples, the spectra exhibited a distinct minimum in the linewidth at the Néel temperature, $T_N \approx 5.5$ K, while the resonance field practically does not change even on cooling below T$_N$. This is indicative of a small magnetic anisotropy $\sim 320$ G in the antiferromagnetic phase. In the film, the magnetic resonance spectrum is split below $T_N$ into several components due to excitation of the magnetostatic modes, corresponding to a non-uniform precession of magnetization. Moreover, the film was observed to degrade over two years. This was manifested by an increase of defects and a change in the domain structure. The saturated magnetization in the film, estimated from the magnetic resonance spectrum, was about 900 emu/cm$^3$ or 5.5 $\mu_B$/unit cell at $T = 3.5$ K.

## 1. Introduction

Ever since the exciting discovery by Katsufuji and Takagi in 2001 that EuTiO$_3$ exhibits a 5% decrease of permittivity below the Neel temperature $T_N = 5.5$ K and a 7% increase of permittivity with the application of a moderate external magnetic field 1.5 T [1], the properties of EuTiO$_3$ have been intensively studied. These large effects were explained by anomalously strong spin-phonon coupling.

EuTiO$_3$ is an incipient ferroelectric like SrTiO$_3$ because its permittivity increases on cooling and saturates below ~20 K due to the softening of a phonon as the temperature is lowered in combination with the quantum fluctuations that occur at low temperature [1,2]. If a magnetic field is applied below $T_N$, the soft mode frequency lowers further, causing permittivity to increase [3]. The room-temperature crystal structure of EuTiO$_3$ is cubic (space group $Pm\bar{3}m$) [4], but near 280 K an antiferrodistortive phase transition to the tetragonal $I4/mcm$ structure occurs [5-8]. The magnetoelectric (ME) coupling is only of third order (i.e., proportional to $H^2E^2$) in EuTiO$_3$ because the crystal and magnetic symmetry exclude a linear ME coupling [9]. Nonetheless, the ME coupling is quite large [9]. Most interestingly, it was found that both ferroelectric and ferromagnetic order can be induced by biaxial strain in thin EuTiO$_3$ films grown on (110) DyScO$_3$ substrates [10,11]. This previously hidden ground state, stabilized by strain, opens a new route for the construction of novel multiferroics in thin film heterostructures with strong ME coupling, a feature not present in bulk materials with the same chemical composition.

Nowadays there are numerous studies on EuTiO$_3$ in the form of bulk ceramics, single crystals, and thin films grown on different substrates (see e.g., [1-8]). In particular, ceramic samples of EuTiO$_3$ have been studied by the electron paramagnetic resonance (EPR) technique down to about 50 K [12]. Spin resonance has been probed using time-domain gigahertz ellipsometry at 4.5 K [13]. To the best of our knowledge, however, EuTiO$_3$ has never been studied using the conventional magnetic resonance technique in the magnetically ordered phase, especially in the form of a thin film.

Since many important questions related to the intrinsic magnetic properties of EuTiO$_3$ are not completely resolved yet and magnetic resonance, in general, provides important information on the microscopic level, such as spin-spin interactions and their thermal fluctuations, magnetic structure, magnetic anisotropy, exchange frequencies and so on, we decided to perform a detailed magnetic resonance investigation of this material for both bulk ceramics and a 100 nm thick epitaxial film deposited on a (110)-oriented DyScO$_3$ substrate over a wide temperature range from 350 K down to 3 K at microwave frequencies of 9.2-9.8 and 34 GHz.

## 2. Experiment

Magnetic resonance (MR) measurements were performed using a Bruker EPR spectrometer operating at 9.2-9.8 GHz (X band) and 34 GHz (Q band) on bulk EuTiO$_3$ ceramics and on a 100 nm thick epitaxial film deposited on a (110)-oriented DyScO$_3$ substrate. For details of the preparation of the bulk EuTiO$_3$ see Refs. [7,14]. The EuTiO$_3$ film was deposited by reactive molecular-beam epitaxy (MBE). Details on the deposition and sample characterization are given elsewhere [10]. The film exhibited 1% biaxial tensile strain. The identical thin film used for these EPR studies was previously measured by far-infrared reflectance spectroscopy [10]; similar, but thinner (~20 nm) films, grown in the same lab by MBE were the subject of detailed magnetic, structural, and second-harmonic studies [10,15]. In particular, strain-induced ferroelectricity was found in this same film below 250 K. Although MOKE measurements showed ferromagnetic order below 4.3 K [10], detail magnetic force microscopy studies of EuTiO$_3$/DyScO$_3$ revealed a coexistence of ferromagnetic and nonferromagnetic states at low temperatures [15].

For MR measurements ceramic samples were ground into a powder with a particle size of 5-10 μm in order to minimize the influence of the demagnetization fields and conductivity on magnetic resonance spectra. In order to accurately measure the intrinsic linewidth near the Néel temperature, $T_N$, an extremely small amount (~ 0.1 mg) of the powder was used for measurements below 50 K. At these low temperatures even in a 100 nm thin film, the signal is strong (it is proportional to the static magnetic susceptibility $\chi_0$).

The magnetometric measurements were performed by means of a Quantum Design Physical Property Measurement System (PPMS) fitted with a vibrating sample magnetometer probe (VSM) and a SQUID magnetometer MPMS-5S. The magnetic susceptibilities were measured under zero-field cooled (ZFC) and field-cooled (FC) protocols, in the temperature range between 2 and 300 K.

## 3. Experimental results and discussion

### 3.1. EPR spectra

Let us first consider the behavior of the EPR spectra of the paramagnetic phase. In this phase we measure the conventional EPR spectra of $Eu^{2+}$ ions, which have electron spin $S = 7/2$. At room temperature, the EPR spectrum of ceramics consists of only one very broad line of nearly Lorentz shape with a peak-to-peak linewidth of 2.24 kOe in the Q band and about 2.90-2.95 kOe in the X band (Figure 1). The linewidth is markedly narrower in the Q band as compared to the linewidth in the X band. This is related to the fact that the spectrum in the X band is also affected by forbidden transitions due to the Zeeman energy ($g\beta H \sim 0.3$ cm$^{-1}$) in this band being comparable to the energy of the dipole interaction between $Eu^{2+}$ spins ($\sim \beta^2 g^2 S^2/r^3 = 0.37$ cm$^{-1}$), where $\beta$ is the Born magneton, $g \approx 2$ is the $g$ factor of $Eu^{2+}$, and $r$ is the distance between the $Eu^{2+}$ ions. The dipole energy is only a small perturbation to the Zeeman energy in the Q band.

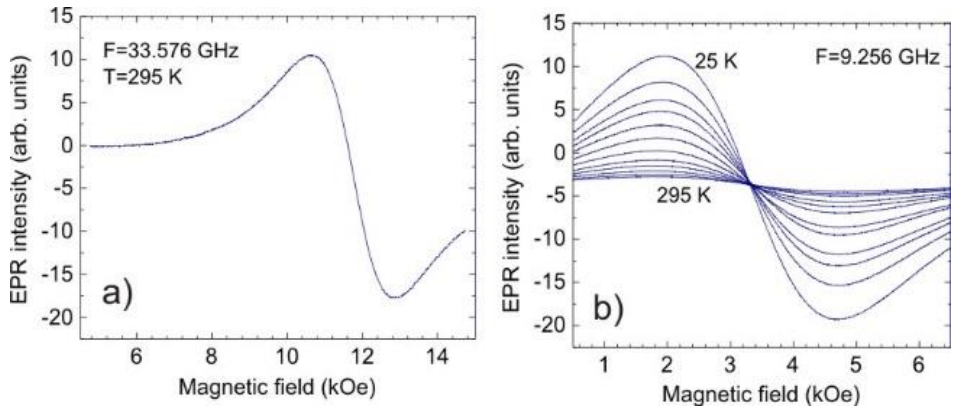

**Figure 1.** (a) Room-temperature EPR spectrum of bulk EuTiO$_3$ ceramics in the Q band. (b) Temperature dependence of the EPR spectra of the same ceramics taken in the X band at temperatures 25-295 K.

On cooling, the intensity of the EPR spectrum increases according to the Curie-Weiss law (Figure 1(b)), but the linewidth practically does not change with temperature down to 200 K (Figure 2(a)). With further temperature lowering, the linewidth starts to decrease.

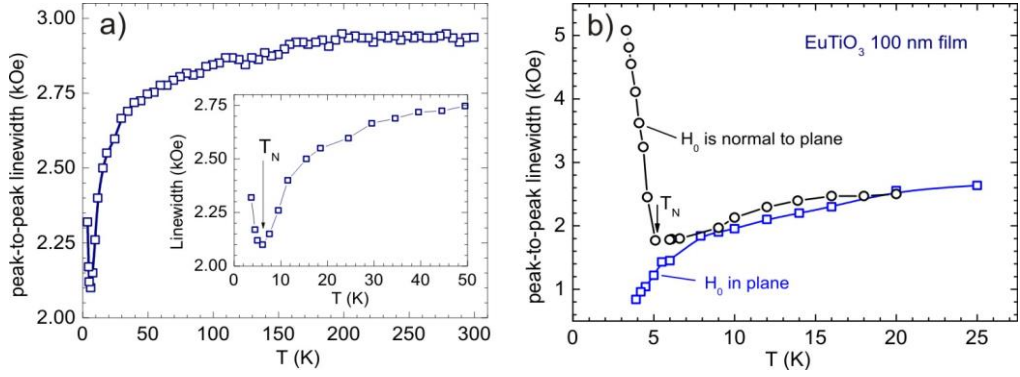

**Figure 2.** Temperature dependence of the EPR linewidth measured in bulk EuTiO$_3$ ceramics (a) and in the 100 nm thick biaxially strained film (b) at two magnetic field directions: normal to the film plane and in the film plane.

The small decrease of the linewidth observed in bulk EuTiO$_3$ down to 50-100 K is comparable to the effect reported in Ref. [12]. Nevertheless, a new sharp narrowing of the line occurs below 30 K (see the inset to Figure 2(a)). Moreover, the linewidth exhibits a distinct minimum at a temperature of 5.5-6 K, where EuTiO$_3$ undergoes the antiferromagnetic (AFM) phase transition [1,16].

The temperature behavior of the linewidth of the EuTiO$_3$ film is shown in Figure 2(b). Due to the small volume of the film, the EPR spectrum could only be measured below 30 K with a satisfactory signal-to-noise ratio. Below $T_N$, a big difference appears between the linewidth measured with the magnetic field direction normal to the film plane and in the film plane. This is related to the magnetic anisotropy in the magnetically ordered phase, as the film is structurally ordered, and the influence of demagnetization field, which is large in the film due to its small thickness. The effect of the demagnetization field is already perceptible in the paramagnetic phase, because the resonance field changes on cooling differently for $H_0$ applied parallel to or perpendicularly to the film plane (Figure 3). Such anisotropic behavior was not observed in the powder from bulk EuTiO$_3$, because the particles had nearly spherical or rectangular shape and random orientations with respect to the external field. The demagnetization field $H_D$ depends on the shape of the sample and its magnetization $M_0$: $H_D = -N_D M_0$. Here, $N_D$ is the sample-shape-dependent demagnetization factor. For instance, it is maximal ($N_D = 4\pi$) if $H_0$ is applied perpendicular to the film plane and $N_D$ is zero when the field is applied in the plane of the films. In spherical particles the demagnetization effect does not depend on field orientation [17].

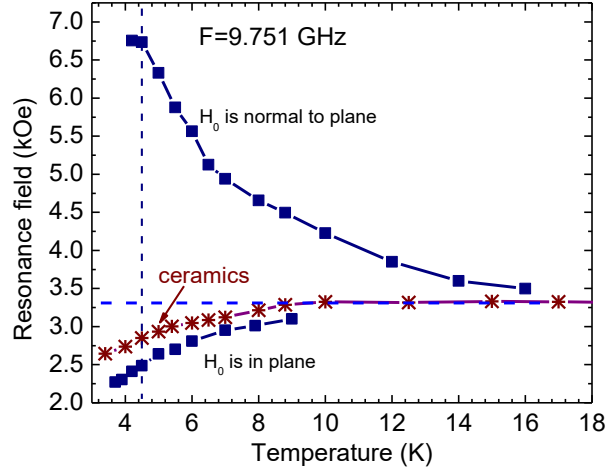

**Figure 3.** Temperature dependence of the resonance field in bulk EuTiO$_3$ ceramics and for two orientations of the EuTiO$_3$ thin film. In the latter case, the magnetic field $H_0$ was applied either perpendicular to or parallel to the plane of the film. The vertical dashed line indicates the temperature $T_N$.

Due to the effect of the demagnetization, the resonance field of the film depends on its orientation with respect to the direction of the magnetic field even in the paramagnetic phase. As an example, the angular dependence of the resonance field in the film taken at 8 K is shown in Figure 4.

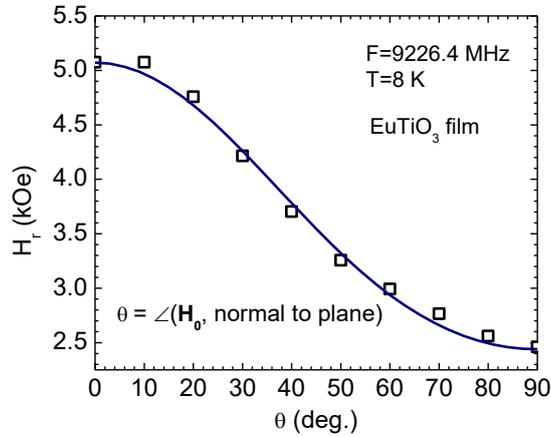

**Figure 4.** Angular dependence of the resonance field measured in the EuTiO$_3$ film at 8 K. $\theta$ is the angle between the magnetic field and the normal to the plane of the film.

The observed angular variation of the resonance field, like for ferromagnetic resonance in thin films, is described by the well-known expression, which in addition to the electron Zeeman term also contains the demagnetization term with $N_D = 4\pi$ [17]:

$$\left(\frac{\omega}{\gamma}\right)^2 = [H_r - 4\pi M \cos 2\theta][H_r - 4\pi M \cos^2 \theta], \qquad (1)$$

where $H_r$ is the resonance field at the microwave frequency $\omega$, $\gamma = g\beta/\hbar$ is the gyromagnetic ratio, and $M$ is the magnetization induced by the external magnetic field. One can see from Figure 4 (solid line) that Eq. (1) describes the angular dependence of the resonance field quite well. Note that the paramagnetic magnetization can be directly measured by the difference between actual resonance field $H_r$ at $\theta = 0$ and the field defined by solely the $g$ factor, i.e., measured at high temperatures, where $M \to 0$:

$$4\pi M = H_r - \frac{\hbar\omega}{g\beta}. \tag{2}$$

*3.2. EPR linewidth*

Let us first analyze the linewidth at $T \gg T_N$, i.e., at room temperature where the crystal structure of EuTiO$_3$ is simple cubic [4,8].

In a paramagnetic cubic phase, the EPR spectrum is mainly determined by the exchange and magnetic dipole interactions. Usually, the spectrum is exchange averaged and narrowed. The large linewidth measured in EuTiO$_3$ is due to the anisotropic spin-spin interaction of the Eu$^{2+}$ ions, which are predominantly of a dipole type. The dipole linewidth is usually calculated by the moment method. For a cubic lattice this gives [18]:

$$M_2 = 5.1(g\beta N)^2 S(S+1), \tag{3}$$

and the peak-to-peak linewidth for the Lorentzian line is equal to

$$\Delta H_d \approx \frac{2}{\sqrt{3}}\sqrt{M_2}. \tag{4}$$

The value $M_2$ in Eqs. (2-3) represents the second moment of the dipole broadening, $N$ is the density of spins per cubic centimeter, $\beta$ is the Bohr magneton, $S = 7/2$ and $g \approx 2$ are the Eu$^{2+}$ spin and $g$ factor, respectively.

The dipole linewidth for the EuTiO$_3$ lattice according to Eq. 4 is $\Delta H_d = 3.8$ kOe. This is larger than the linewidth ~2.9 kOe measured at room temperature. Therefore, the Eu$^{2+}$ EPR spectrum in EuTiO$_3$ is partly exchange narrowed due to the rapid exchange between spins. The exchange-narrowed linewidth can be estimated from the well-known relation [18]:

$$\Delta H_\infty \approx \frac{\gamma(\Delta H_d^2 + H_{cf}^2)}{\omega_e}, \tag{5}$$

where $\omega_e$ is the exchange frequency and $\gamma H_{cf}$ is the zero field splitting of the Eu$^{2+}$ spin levels. The zero field splitting term (crystal field term) is usually much smaller than the magnetic dipole linewidth in materials with cubic crystal structure. As a good estimation of the order-of-magnitude value of the crystal field, well known data for Eu$^{2+}$ in SrTiO$_3$ can be used. Both materials have practically the same lattice constant, a = 3.905 Å at room temperature. The largest 4-th rank term $b_4^0$ of the crystal field in SrTiO$_3$ is only 106×10$^{-4}$ cm$^{-1}$ and the tetragonal-symmetry term $b_2^0$ related to oxygen octahedral rotation is only 10×10$^{-4}$ cm$^{-1}$ at 4.2 K [19]. Therefore, we can neglect the crystal field term.

The exchange frequency is usually calculated from the exchange energies $J_{jj'}$ [18,20,21]:

$$\hbar^2 \omega_e^2 = \frac{2}{3} S(S+1) \sum_{j'} J_{jj'}^2 \qquad (6)$$

where the two indices $j,j'$ run over all nearest- and next-to-nearest neighbors. The nearest-neighbor and the next-nearest-neighbor exchange interaction energies in EuTiO$_3$ are very small: $2J_1 = -0.037$ K and $2J_2 = 0.069$ K [1], respectively. Larger values of the exchange interaction ($J_1 = -0.021$ K, $J_2 = 0.040$ K) have also been reported in a much older paper from 1966 [16]. This leads to the exchange frequency $\omega_e/2\pi$ = 15-15.5 GHz and the exchange field $H_e = \omega_e/\gamma$ between 5-6 kOe. This value agrees well with the critical field of the spin-flip transition, $H_{C2} = 2H_e \approx 10$ kOe, determined from magnetic measurements (see, section 3.3 and Refs. 9,16,22). Note that a much higher exchange frequency $\omega_e/2\pi \approx 41.25$ GHz was reported in [12] where, however, the nearest-neighbor exchange interaction constant $J_1 \approx 0.26$ K was used in the estimation of the exchange frequency.

*3.3. Temperature dependence of the linewidth and spectra intensity*

The EPR linewidth of EuTiO$_3$ exhibits a quite unusual temperature behavior compared to other antiferromagnets. First, the linewidth critically narrows on cooling towards the Néel temperature and second, the spectrum does not disappear at $T < T_N$, indicating that the magnetic anisotropy is small as it is in AFM phases like cubic KMnO$_3$ [23] or RbMnO$_3$ [24]. In typical AFM materials the EPR spectra usually critically broaden on approaching the Néel temperature and the AFM resonance cannot be easy observed below $T_N$ [25-27] due to the appearance of a large (up to tens of cm$^{-1}$) forbidden gap in the spin excitations. Quantitatively, the decrease of the linewidth can be understood from the general expression used for the description of the linewidth of a paramagnetic phase of a magnetic material [25,28]:

$$\Delta H = (\Delta H)_\infty \frac{\Gamma(T)}{T \chi(T)}, \qquad (7)$$

where $\chi$ is the static magnetic susceptibility and $\Delta H_\infty$ is the temperature-independent linewidth determined by the dipole interactions in Eq. (5). For EuTiO$_3$, $\Delta H_\infty$ can be taken as the linewidth at room temperature. The function $\Gamma(T)$ accounts for the temperature dependence of the spin-spin correlations. This function is temperature independent at $T >> T_N$, but it increases significantly in the vicinity of the AFM phase transition as the spin fluctuations slow down. It can be approximated as [28]:

$$\Gamma(T) = A\left(\frac{T-T_N}{T_N}\right)^{-\delta}, \qquad (8)$$

where the exponent $\delta$ can vary over the range 0.5-2 and $A$ is a scaling factor which depends on the specific lattice parameters and the nature of the anisotropic spin-spin interactions. One can see from Eq. (7) that the linewidth mainly depends on the competition between the two contributions: the product of the magnetic susceptibility $\chi$ and temperature, which reduces the linewidth, and the function $\Gamma(T)$, which broadens the linewidth. In order to prove the existence of anisotropic spin-spin interactions, we plot the product $\Delta HT\chi$ against the reduced temperature $(T-T_N)/T_N$ on a log-log scale in Figure 5. The values of $\chi(T)$ were extrapolated by the Curie-Weiss law with the positive Curie-Weiss temperature $\theta \approx 3.2\text{-}3.8$ K [1,16]. Namely, due to the positive Curie-Weiss temperature $T_\theta$ in EuTiO$_3$, the term $(T\chi)^{-1}$ in Eq. (7) diverges near the Néel temperature, effectively decreasing the EPR linewidth.

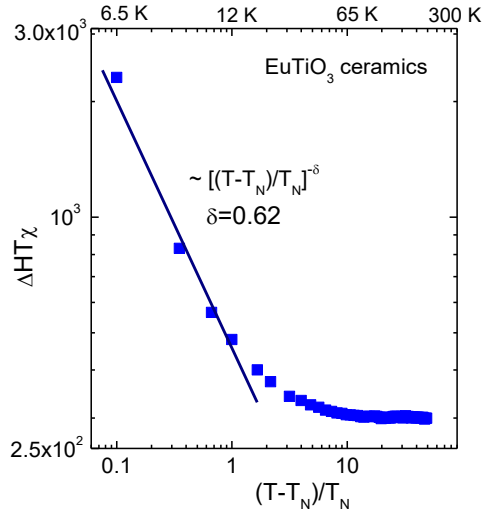

**Figure 5.** Log-log plot of the product $\Delta HT\chi$ as a function of the reduced temperature $(T-T_N)/T_N$.

On the other hand, one can see from the plot in Figure 5 that $\Delta HT\chi$ varies as $[(T-T_N)/T_N]^{-0.62}$ in the temperature range $0.1 < (T-T_N)/T_N < 1$. Further, up to 6 K above $T_N$, the non-zero contribution to the linewidth comes from the anisotropic spin-spin interactions, which is predominantly of magnetic dipole origin with some contribution from crystal fields. For higher temperatures, the product $\Delta HT\chi$ is temperature independent and the linewidth changes as $(T\chi)^{-1}$, i.e., the function $\Gamma(T)$ in Eq. (7) is temperature independent.

The slope of the straight line in Figure 5 gives the parameter $\delta$, which is directly related to the critical exponent of the spin correlation length $\xi$ via the approximate relations (see, e.g., Ref. [29] and references therein):

$$\begin{aligned}\delta &\sim (1.5-2.5)\nu \\ \xi &\sim (T-T_N)^{-\nu},\end{aligned} \qquad (9)$$

where the critical exponent $\nu \approx 0.5$.

The temperature variation of the EPR spectrum integral intensity $I(T)$ provides information on the type of the $Eu^{2+}$-$Eu^{2+}$ magnetic interaction as $I(T)$ is proportional to the static magnetic susceptibility $\chi(T)$. Such data is shown on the $EuTiO_3$ thin film in Figure 6(a). Note that these EPR data present the contribution of only the $Eu^{2+}$ ions without the influence of the paramagnetic substrate, $DyScO_3$. This is in contrast to the direct magnetic susceptibility measurements, where the substrate contribution dominates the total magnetization as can be seen from Figure 6(b). In the magnetic measurements the contribution of the film magnetization to the total magnetization is only reflected by the weak bump at a temperature of 5 K. The anomaly at a temperature 3.1 K is due to the AFM phase transition of the $DyScO_3$ substrate [30]. On the other hand, the magnetic field used in the EPR measurement is obviously higher than the critical field of the spin-flop transition, which is in the range of 1-2 kOe in bulk $EuTiO_3$ depending on temperature [9,22]. Therefore, the anomaly near $T_N$ of the film (~ 5 K) is smeared out but is well manifested in EPR linewidth (Figure 2).

For comparison, the temperature dependence of the inverse magnetic susceptibility obtained from magnetometric measurements on a bulk $EuTiO_3$ ceramic sample at three characteristic values of magnetic field is shown in Figure 6(c). One can see that the inverse susceptibility has a distinct minimum at the Néel temperature in a low field of 100 Oe. This minimum diminishes in a field of 1 kOe and completely disappears when the field increases above the critical value of the spin-flip transition and the susceptibility behaves like in a paramagnet (see the data at 10 kOe).

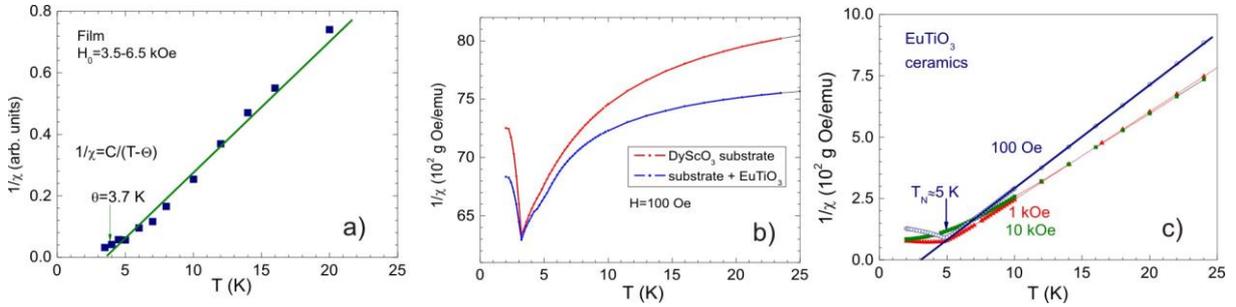

**Figure 6.** The temperature dependence of the inverse magnetic susceptibility measured by (a) EPR and (b) SQUID magnetometry on the $EuTiO_3$ film and (c) the inverse of the susceptibility of bulk $EuTiO_3$ obtained from magnetic measurements at three magnetic fields: 100 Oe, 1 kOe, and 10 kOe showing the Curie-Weiss behavior with Curie temperatures of 3.7 and 3.2 K for the film and bulk $EuTiO_3$, respectively. The anomaly at the Néel temperature diminishes in the 10 kOe field due to the spin-flop and spin-flip transitions. Since the susceptibility of the $DyScO_3$ substrate under the $EuTiO_3$ film is strongly dependent on crystal orientation, the two curves in the graph (b) do not exactly coincide.

*3.4. Magnetic resonance spectra at $T < T_N$*

As was mentioned above, the magnetic resonance spectrum does not disappear in the AFM phase below $T_N$. Rather, it transforms into the spectrum of AFM resonance. There is no visible change in the spectral line shape at the Néel temperature $T_N = 5.5$ K in bulk EuTiO$_3$ (Figure 7(a)).

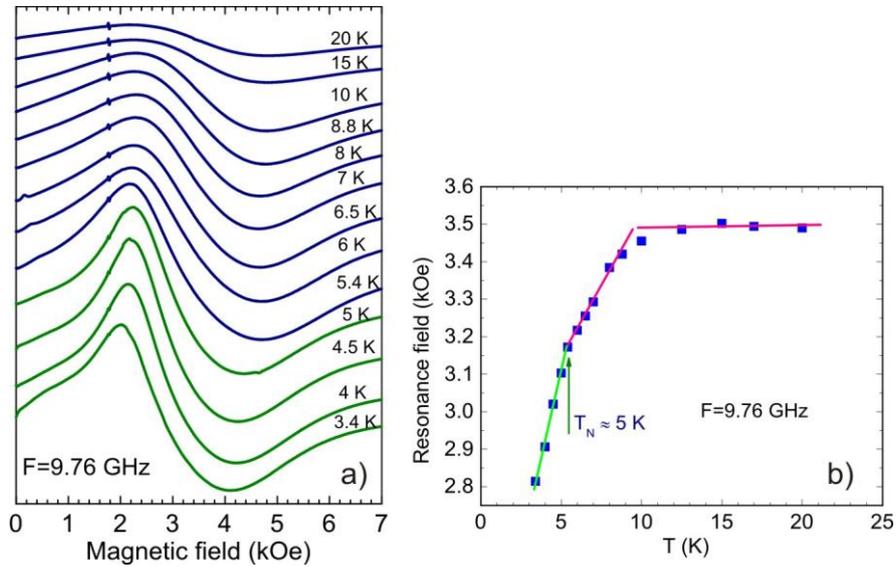

**Figure 7.** (a) Magnetic resonance spectra and (b) temperature dependence of the resonance field in bulk EuTiO$_3$ in the vicinity of the AFM phase transition at $T_N \approx 5.5$ K.

Nevertheless, the resonance field already starts to shift downward in the paramagnetic phase on cooling below 10 K due to the increase of the demagnetization field (Figure 7(b)). This shift becomes steeper at the phase transition as the magnetic anisotropy appears below $T_N$.

Larger changes are visible in the EPR spectra of the film. If a magnetic field is applied perpendicular to the film plane, the spectral line in the paramagnetic phase shifts towards higher magnetic fields as the temperature decreases down to the Néel temperature due to the increase of the demagnetization field on cooling (Figures 8 and 3).

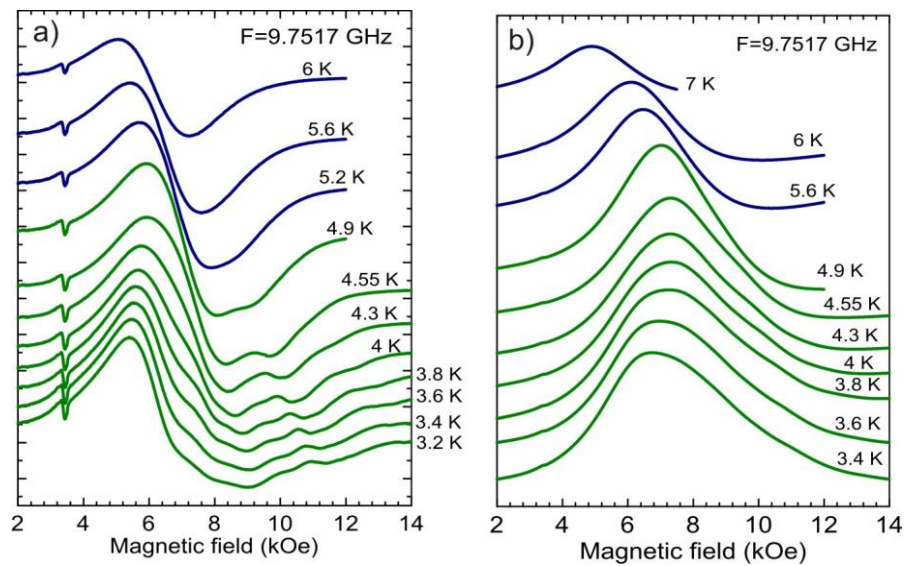

**Figure 8.** Temperature dependence of the magnetic resonance spectra of the EuTiO$_3$ film in the vicinity of and below the AFM phase transition temperature with $T_N \approx 5.1$ K. The first derivatives of the spectra are shown in panel (a); the absorption spectra are shown in panel (b). $H_0$ is applied perpendicular to the plane of the film.

Below the Néel temperature, the spectral line in the film splits into several components due to the formation of magnetic order with magnetic anisotropy and domain structure. As can be seen from Figure 8, the strongest spectral component shifts below $T_N$ towards lower magnetic fields while other components lower in intensity and strongly shift towards higher magnetic fields. The splitting in the spectrum increases as the temperature is lowered. The behavior of the main spectral component is clearly seen in the absorption spectra shown in Figure 8(b). The resonance field in the AFM phase depends on many parameters including the sublattice magnetization as well as the external, exchange, and anisotropy fields [17]. Note that under applied magnetic fields, the AFM phase corresponds to the spin-flop state, because the critical field of the spin-flop $H_C \approx \sqrt{2H_e H_A} \approx$ 1.5-1.8 kOe at $T = 3.5$ K is much lower than the applied field at resonance due to the low exchange $H_e$ and anisotropy $H_A$ fields. Moreover, when the external field increases, the spin system undergoes a transition from the spin-flop to the spin-flip states as the critical field of the spin-flip transition $H_{C2} = 2H_e$ is only about 10 kOe. Therefore, a quantitative description of these spectra is a quite complex task, which is beyond the scope of this paper.

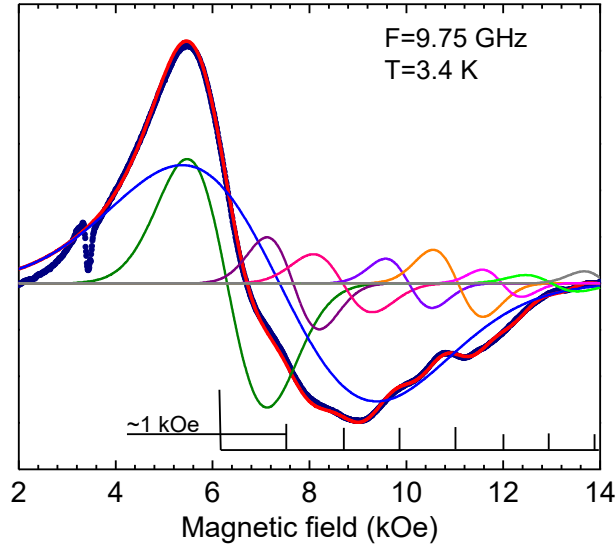

**Figure 9.** Decomposition of the complex spectral line in the EuTiO$_3$ film measured at 3.4 K into individual components separated by a distance of approx. 1000 Oe. A narrow line near 3.3 kOe is an experimental artifact.

Note that the low-temperature spectrum in the film can be decomposed into a number of components separated by approximately the same distance of 1000 Oe as is shown in Figure 9. This suggests that either the magnetostatic or standing-spin-wave AFM modes are excited. They correspond to the non-uniform precession of magnetization.

An estimation of the distance between the standing-spin-wave modes [31,32] gives a value of only ~0.4$m$ Oe, where $m$ is an integer. This is related to the fact that the separation between the spin wave modes is proportional to the exchange energy, which is very small in EuTiO$_3$. Therefore, any resonances due to these modes are completely unresolvable from the main mode.

The theory of magnetostatic mode excitations in thin plates of a ferro- or ferri-magnet was considered in Refs. [31,33]. Because the magnetization in the EuTiO$_3$ film at the resonance fields is practically saturated, this theory with some precaution can be applied to the spectrum in Figure 9. In particular, theory predicts the right edge of the spectrum to be at $H_r = \omega/\gamma + 4\pi M \approx 14$ kOe in agreement with our experiment. Calculation of the resonance fields of the magnetostatic modes can be performed only numerically. The results of such calculations are presented in [31] for EuS. They predict an increase of the magnetic field separation between the modes from 200-400 Oe for a field orientation in-the-plane of the plate and up to 1000-1400 Oe for an out-of-plane oriented field. This completely agrees with our observation on the EuTiO$_3$ film: when the magnetic field is applied in the plane of the film, the magnetostatic modes are not resolved, but they become visible and the separation between modes increases to 1000 Oe, when the magnetic field is applied normal to the film plane.

Note that SQUID [11], polarized neutron reflectometry [11], and MFM studies [15] of the 20 nm thick EuTiO$_3$ film (grown on DyScO$_3$) revealed magnetic anisotropy in the film plane in the stress-induced ferromagnetic phase and even the possible coexistence of magnetic and paramagnetic states, respectively [15]. In the 100 nm thick film, however, which obviously undergoes a normal AFM phase transition, the anomaly at the Néel temperature detected by the SQUID magnetometer is so weak (see Figure 6(b)) that the magnetic anisotropy can hardly be resolved. Nevertheless, the magnetic anisotropy in the film plane is clearly seen in the magnetic resonance spectra (Figure 10). When the applied field is rotating in the film plane, the effect of demagnetization is angle independent (see Eq. 1) and all changes in resonance field relate only to the magnetic anisotropy. The magnetic anisotropy shows $180^0$ periodicity that obviously agrees with the easy-axis of the AFM structure, where the easy axis lies in the film plane [17,22].

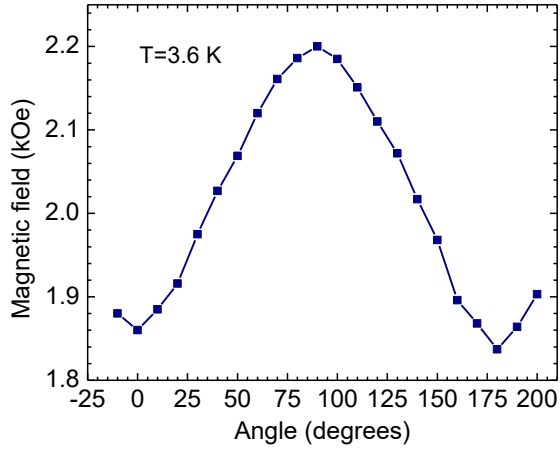

**Figure 10.** Angular dependence of the resonance field measured in the plane of the film at 3.6 K showing $180^0$ periodicity.

We would like to stress that the low-temperature spectra of the film change over time. Figure 11 shows spectra measured on the $EuTiO_3$ film two years after it was grown. One can see that the spectral density is now distributed continuously between minimal and maximal resonance fields and the magnetostatic modes are no longer resolved, indicating that the magnetic domain structure has changed in the film over time. Structural imperfections are expected to increase with time in strained films and $Eu^{2+}$ is expected to oxidize to $Eu^{3+}$ over time in contact with air. Such degradation is likely accelerated by the extensive measurements made on this film in different measurements with temperature cycles between 2 and 300 K. From the right edge of the spectrum, the magnetization in the film can be estimated as 900 G or emu/cm$^3$, which is smaller than its saturated value of 1100 G in bulk $EuTiO_3$ [22]. In contrast, fresh unstrained $EuTiO_3$ films produced in the same manner as this one exhibit expected values of saturated magnetization, 6.7 +/- 0.5 Bohr magnetons per europium atom [34]. Over time, the strained $EuTiO_3$ film grown on (110) $DyScO_3$ degrades with exposure to air and temperature cycling, as has been noted in a prior study [11].

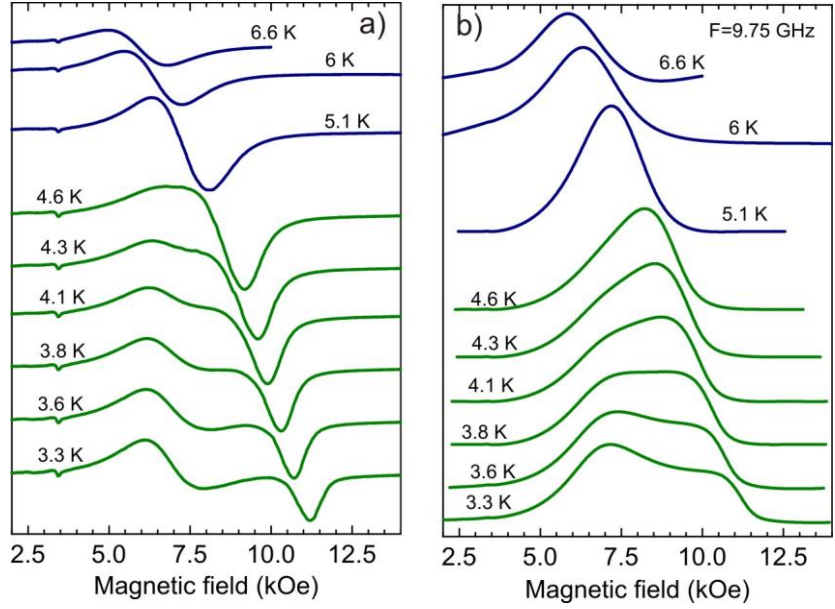

**Figure 11.** Temperature dependence of the magnetic resonance spectra in the vicinity of and below $T_N$ in a film two years after it was synthesized. The first derivatives of the spectra are shown in panel (a); the absorption spectra are shown in panel (b).

## 4. Conclusion

Magnetic resonance spectra were measured on EuTiO$_3$ in both bulk and thin films form over a wide temperature interval from 350 down to 3 K at microwave frequencies of 9.2-9.8 and 34 GHz, i.e., in the X and Q bands, respectively. In the paramagnetic phase, the spectrum consists of one very broad line with a peak-to-peak width of 2.90-2.95 kOe in the X band at room temperature. On cooling, the spectrum increases in intensity according to the Curie-Weiss law with positive Curie-Weiss temperature $T_\Theta$ = 3.2-3.7 K. With decreasing temperature, the linewidth practically does not change down to about 150 K. On further cooling, however, it sharply decreases and has a distinct minimum at the temperature of the AFM phase transition. From the high-temperature linewidth we estimate the exchange frequency to be $\omega_e/2\pi$ = 15-15.5 GHz and the exchange field to be $H_e = \omega_e/\gamma$, between 5-6 kOe.

The temperature dependence of the linewidth is explained by spin-spin correlations, which exhibit critical behavior as the Néel temperature is approached. From a fit to the experimental data, the critical exponent $\nu \approx 0.5$ of the spin correlation length $\xi \sim (T - T_N)^{-\nu}$ is estimated.

The magnetic resonance spectrum does not disappear in the AFM phase. It transforms into the spectrum of the AFM resonance visible in the X-band. There is no visible change in the spectral line at the Néel temperature, indicating that the magnetic anisotropy is small enough, about 320 G, that it corresponds to the critical field of the spin-flop transition ~1800 Oe at $T$=4 K. Nevertheless, the resonance field already starts to shift in the paramagnetic phase due to the increased contribution from the demagnetization field. The effect of the demagnetization field is especially strong in the EuTiO$_3$ film due to its shape and the resonance field essentially depends on the EuTiO$_3$ film orientation with respect to the magnetic field direction.

The spectral line in the EuTiO$_3$ film splits into several components below the Néel temperature. We have shown that this multi-component spectrum can be explained in terms of the excitation of the magnetostatic modes. Despite the small thickness of the EuTiO$_3$ film ($\approx$100 nm), resonances from the standing-wave spin waves were not resolved due to the very small exchange energy of the Eu$^{2+}$ spins in the EuTiO$_3$ lattice. Overall, the properties of this EuTiO$_3$ film are similar to those of bulk EuTiO$_3$ excepting some ageing effects.

It is worth noting that the magnetic resonance technique has advantages over the SQUID technique for the characterization of EuTiO$_3$ thin films. Being even more sensitive, it allows the intrinsic characteristics of EuTiO$_3$ films to be measured without being dominated by the contributions from the paramagnetic substrate. Moreover, the sublattice magnetization and magnetic anisotropy can be determined without knowledge of the volume or the mass of the sample.

**Acknowledgment**

The support of the Czech Science Foundation under projects No. 13-11473S, 15-06390S and 15-08389S is gratefully acknowledged, as well as support of projects SAFMAT LM2015088, LO1409 and MEYS under the project CEITEC 2020 (LG1601). J.H.L. and D.G.S. acknowledge support from the Center for Nanoscale Science, a National Science Foundation center through Grant number DMR-1420620.